# Numerical Simulation of Shock Wave Propagation Over a Dense Particle Layer Using the Baer-Nunziato Model

P. Utkin[1,*], P. Chuprov[2]

[1] Harbin Institute of Technology,
No. 92 West Dazhi Street, Nan Gang District, Harbin, Heilongjiang Province, 150001, China

[2] Institute for Computer Aided Design of the Russian Academy of Sciences,
19/18 2nd Brestskaya, Moscow, 123056, Russia

Corresponding author, e-mail: utkin@hit.edu.cn

**Abstract** The present study examines the possibility of numerical simulation of a strong shock wave propagating over the surface of a dense layer of particles poured onto an impermeable wall using the Baer-Nunziato two-phase flow model. The setting of the problem follows the full-scale experiment. The mathematical model is based on a two-dimensional system of Baer-Nunziato equations and takes into account intergranular stresses arising in the solid phase of particles. The computational algorithm is based on the Harten-Lax-van Leer-Contact (HLLC) method with a pressure relaxation procedure. The developed algorithm proved to be efficient for two-phase problems with explicit interfacial boundaries and strong shock waves. These issues are typical of problems arising from the interaction of a shock wave with a bed or a layer of particles. A comparison with the simulations and full-scale experiments of other authors is carried out. A reasonable agreement with the experiment is obtained for the angles of the transmitted compaction wave and granular contact, including their dependency on the intensity of the propagating shock wave. The granular contact angle increases with the incident shock wave Mach number, while the transmitted compaction wave angle decreases. An explanation is given of the phenomenon of the decrease in thickness of the compacted region in the layer with the increase in intensity of the propagating shock wave. The main reason is that the maximal value of the particle volume fraction in the plug of compacted particles in the layer rises with the increase in shock wave intensity.

## 1. Introduction

A dust explosion is one of the most dangerous scenarios for the development of an emergency situation in industries associated with fine chemically reacting powders. Examples include not only coal mines but also the production of medicines and food products. In the latter case, the danger is, for example, starch particles and powdered sugar. In recent years, aluminum powder explosion accidents have become one of the main types of dust explosion accidents, causing serious consequences[1]. Such accidents can occur in industries related to the polishing of metal products. For example, as noted by Wang et al.[1], the explosion accident in Guangdong, China, on November 20, 2012, was caused by an electrostatic spark igniting the deposited aluminum powder layer in a ventilation duct. On August 2, 2014, the explosion accident in Jiangsu, China (at least 68 people were killed) was caused by the explosion of aluminum powder due to complex environmental conditions of high temperature and high humidity.

Let's describe schematically the mechanism of a layered dust explosion using the example of an explosion in a coal mine[2]. Generally, the explosion begins near the closed end of the mine gallery, where methane accumulation is possible. The burning of the coal dust particle occurs in the following way. The matter that burns is the volatiles, which are combustible gases released from the heated particle. After some time, generally tens of milliseconds, the flame front covering the whole cross section of the gallery is formed. The moving flame front pushes the air ahead of it, creating the shock wave behind which the gas flows. The hot flame front radiates heat to the mixture ahead of it. The leading shock wave propagates over the coal dust layer; the deposited dust is lifted before the flame front arrives and subsequently undergoes burning. This results in the heating of the nearby unburned particles, which in turn release volatiles to further fuel the combustion process. In such a way, a coal dust explosion can propagate in the mine gallery until the whole mass of the dust is burned or propagation is interrupted by protection means. So, the process of shock wave propagation over the dust layer is an important stage of the layered dust explosion's development.

The process of dust explosion development has been simulated in several recent papers. Shimura and Matsuo[3] used a computational fluid dynamics – discrete element method model of compressible flow for the numerical study of the flame structure during shock-wave-induced layered coal-dust combustion. They found that the particle dispersion heights were higher than those previously predicted by the Eulerian-Eulerian approach for the same statement of the problem. In turn, it led to the attenuation of the compression wave from the reaction front and the slowing of the leading shock wave. In those simulations, Song and Zhang[4] separated out four possible regimes of coal dust combustion development depending on the thickness of the layer and the size of particles, namely, non-ignition, halfway quenching, slow combustion, and explosion. The boundaries between these regimes were identified in a quantitative manner. Guhathakurta and Houim[5] performed numerical simulations to study the impact of thermal radiation and particle diameter on layered coal-dust explosion. The results show that the impact of radiation on the dust explosion is situation specific. Radiation can promote, inhibit, or have little impact on the explosion depending on the particle diameter.

Although the authors of the earlier alluded papers contributed to the study of different aspects of dust explosion, wave dynamics inside the particle layer during the shock wave propagation over the layer are much less studied in comparison with the processes that occur after the dispersion of particles from the layer. We can note recent works related to the study of wave processes inside dense layers of granular media when a shock wave propagates over them. Khomik et al.[6] measured the overpressure and impulse of the compression phase transmitted to a rigid wall through the layer of dispersed material during the propagation of a spherical shock along it in experiments using a conical shock tube. The layers of sand of different fineness and thickness were investigated by varying the intensity of the shock wave. Sugiyama et al.[7] numerically investigated the interaction of a shock wave and particles filling inside of a straight tube to understand the mitigation mechanism of a blast wave outside the tube. It was shown that the heat transfer from the air to the particle layer was the dominant factor in mitigating the blast wave. To estimate this effect, a novel index was proposed. A Baer-Nunziato-type[8], two-phase flow model was utilized to model the granular medium. The convective and non-conservative terms were discretized using an HLLC-type Riemann solver[9]. A dust lifting process was simulated by Le et al.[10] using the original compressible multiphase particle-in-cell method. Special attention was paid to the wave dynamics inside the layer. The compression wave in the solid phase and the corresponding reflected wave due to the interaction of a compression wave with the wall were resolved. It was shown that when the reflected wave reached the surface of the layer, huge collision forces acted on the particles and significantly promoted the initial lifting of the particles.

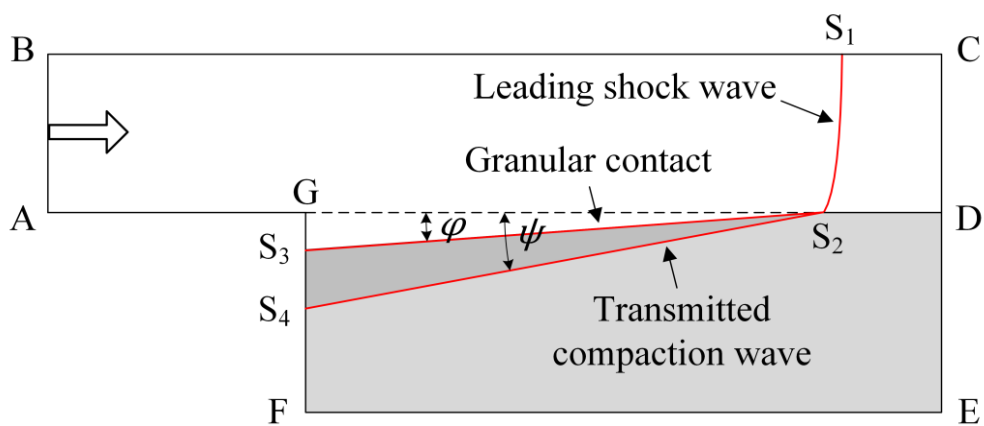

**Fig. 1.** The schematic of the computational area and the sketch of the flow being realized.

In contrast to the problem of the normal incidence of a shock wave on a layer of particles, which has been studied extensively for many years both experimentally and numerically, see the references in Ref. 11, earlier alluded papers[6, 7, 10] were devoted to the study of the multidimensional effects of a shock wave – particle layer interaction. The work of Fan et al.[12] (as well as the earlier mentioned paper of Khomik et al.[6]) contains rare experimental data that can be used to verify multidimensional two-phase models for shock wave – particle layer interaction. In the experiments of Fan et al.[12], a shock wave of varying intensity propagated over a layer of starch particles. Under the action of an incident shock wave, the transmitted compaction wave, induced inside the granular material, formed an oblique angle $\psi$ (see Fig. 1). The dusty mass surface behind the shock wave sloped down with a turning angle $\varphi$. Both angles were visualized and measured in the experiments, depending on the intensity of the incident shock wave, by means of shadowgraphs and pulsed X-ray radiography with trace particles added. Fan et al. found that the compressed region in the dusty bulk mass behind the incident shock wave was reduced with the increasing propagation velocity of the incident shock wave. The reduction of the compressed region was due to the increase in angle $\varphi$ and decrease in angle $\psi$ with the increase in incident shock intensity. The dependencies obtained were not explained, although Fan et al. carried out the numerical simulations of their experiments. The mathematical model was based on the kinetic theory of granular media[13].

The experiments of Fan et al.[12] were simulated by Khmel and Fedorov[14] and Houim and Oran[15]. In both works, the mathematical model was also based on the kinetic theory of granular media[13]. Khmel and Fedorov[14] considered shock waves of different intensities. They got the correct dynamics for the $\varphi$ and $\psi$ angles with the variation of the intensity of the propagating shock wave but didn't address any discussion of this phenomenon. Houim and Oran[15] only reported the reasonable agreement between experimental and computed values of the angles for one run.

The goal of the current work is to study the possibility of modeling the passage of a shock wave over a layer of particles using the Baer-Nunziato model. The setting of the problem, resembling the experiments of Fan et al.[12], was used as a reference. Our goal is to obtain the correct dependence of the angles on the intensity of the passing wave as well as to explain such dynamics within the framework of the mathematical model used.

This paper is organized as follows: In Section 2, the physical statement of the problem is presented. Section 3 outlines the defining system of equations. In Section 3, the numerical algorithm used in this study is presented in detail. In Section 4, the problem of a shock wave – particle cloud interaction[16] is examined for verification purposes. Section 5 covers simulations of the experiment of Fan et al.[12] on shock wave propagation over the dense particle layer. Section 5.1 is devoted to the methodological issues of defining the angles of the transmitted compaction wave and granular contact. In Section 5.2, the dependence of these angles on the intensity of the propagating shock wave is studied and explained. The conclusions are drawn in the final Section 6.

## 2. Statement of the Problem

Fig. 1 illustrates the setting of the problem and the scheme of the flow being realized. The computational domain ABCDEFG corresponds to a flat channel ABCD with a cavity DEFG filled with starch particles. At the initial instant, the channel is filled with air under normal conditions. The geometric characteristics of the computational domain are the following:

$$AB = 0.02 \text{ m}, AG = 0.02 \text{ m}, BC = 0.17 \text{ m}, DE = 0.02 \text{ m}.$$

In experiments, the distribution of particle diameters and material densities were measured. Fan et al.[12] provided a mean particle diameter of $d = 15$ μm and a true material density of $\bar{\rho}_0 = 1100$ kg/m$^3$. The volume fraction of particles in the layer can also be restored from the experimental data. It was equal to $\bar{\alpha}_0 = 0.47$. So, in our simulations, we consider particles to be spherical, rigid, and monosized with the aforementioned diameter and true density. As is shown by Lai et al.[17], for instance, the effect of polydispersity could be important for the problem of shock wave – dust layer interaction. In particular, lifting dynamics for small and large particles differ. Apparently, for the processes inside the layer, some separation of particles can occur. This issue may be the subject of further study.

The initial level of the upper boundary of the layer is the line GD. Inflow conditions are imposed at the boundary AB. Parameters behind the shock wave with Mach number $M$ are used as inflow values. On the remaining boundaries, slip wall conditions are set. Thus, the shock wave enters the computational domain from the left, passing first over a solid ledge, after which the ledge is replaced by a layer of particles. The line $S_1S_2$ schematically shows the shock wave front, which is bent due to the deceleration of the shock wave part adjacent to the layer. The straight line $S_2S_3$ corresponds to the upper boundary of the layer, which will be lower than the initial level GD as a result of the shock wave passage. The area $S_2S_3S_4$ corresponds to the area of compaction of particles in the layer up to a volume fraction exceeding the value in the original layer.

## 2. Mathematical Model

The mathematical model is based on a two-dimensional system of Baer-Nunziato equations[8]. This model is a cornerstone in two-phase compressible modeling[18]. Each phase is treated as a compressible fluid in local thermodynamic equilibrium, but the mixture is allowed to be in nonequilibrium across the interface. We also note significant progress in the development of Euler-Lagrange-type models for describing the flows of dense two-phase media[3, 4, 19, 20]. However, it is well known that such types of models are less suitable for describing processes that are mainly determined by the interaction of closely spaced particles in comparison with Baer-Nunziato-type models. A number of improvements and further developments of the original Baer-Nunziato system of equations were proposed and justified in papers by R. Saurel[21], J. Bdzil[22], and A. Kapila[23] with co-authors. However, most of these works relate to the problems of the deflagration-to-detonation transition in heterogeneous explosives. Not many works address the issues of the application of Baer-Nunziato-type models to particulate two-phase flows. We utilized the direct generalization of the model used in our previous paper[11] for the two-dimensional case. The two-dimensional model in use is very close to that in the paper[7]. The defining system of equations has the following form:

$$\mathbf{u}_t + \mathbf{f}_x(\mathbf{u}) + \mathbf{g}_y(\mathbf{u}) = \mathbf{h}(\mathbf{u})\bar{\alpha}_x + \mathbf{i}(\mathbf{u})\bar{\alpha}_y + \mathbf{p} + \mathbf{s}, \qquad (1)$$

$$\mathbf{u} = \begin{bmatrix} \bar{\alpha} \\ \bar{\alpha}\bar{\rho} \\ \bar{\alpha}\bar{\rho}\bar{u} \\ \bar{\alpha}\bar{\rho}\bar{v} \\ \bar{\alpha}\bar{\rho}\bar{E} \\ \alpha\rho \\ \alpha\rho u \\ \alpha\rho v \\ \alpha\rho E \end{bmatrix}, \quad \mathbf{f} = \begin{bmatrix} 0 \\ \bar{\alpha}\bar{\rho}\bar{u} \\ \bar{\alpha}(\bar{\rho}\bar{u}^2 + \bar{p}) \\ \bar{\alpha}\bar{\rho}\bar{u}\bar{v} \\ \bar{\alpha}\bar{u}(\bar{\rho}\bar{E} + \bar{p}) \\ \alpha\rho u \\ \alpha(\rho u^2 + p) \\ \alpha\rho uv \\ \alpha u(\rho E + p) \end{bmatrix}, \quad \mathbf{g} = \begin{bmatrix} 0 \\ \bar{\alpha}\bar{\rho}\bar{v} \\ \bar{\alpha}\bar{\rho}\bar{u}\bar{v} \\ \bar{\alpha}(\bar{\rho}\bar{v}^2 + \bar{p}) \\ \bar{\alpha}\bar{v}(\bar{\rho}\bar{E} + \bar{p}) \\ \alpha\rho v \\ \alpha\rho uv \\ \alpha(\rho v^2 + p) \\ \alpha v(\rho E + p) \end{bmatrix}, \qquad (2)$$

$$\mathbf{h} = \begin{bmatrix} -\tilde{u} \\ 0 \\ \tilde{p} \\ 0 \\ \tilde{p}\tilde{u} \\ 0 \\ -\tilde{p} \\ 0 \\ -\tilde{p}\tilde{u} \end{bmatrix}, \quad \mathbf{i} = \begin{bmatrix} -\tilde{v} \\ 0 \\ 0 \\ \tilde{p} \\ \tilde{p}\tilde{v} \\ 0 \\ 0 \\ -\tilde{p} \\ -\tilde{p}\tilde{v} \end{bmatrix}, \quad \mathbf{p} = \begin{bmatrix} F \\ 0 \\ 0 \\ 0 \\ -\tilde{p}F \\ 0 \\ 0 \\ 0 \\ \tilde{p}F \end{bmatrix}, \quad \mathbf{s} = \begin{bmatrix} 0 \\ 0 \\ M_x \\ M_y \\ I \\ 0 \\ -M_x \\ -M_y \\ -I \end{bmatrix}, \qquad (3)$$

$$\alpha + \bar{\alpha} = 1, \qquad (4)$$

$$\bar{E} = \frac{\bar{u}^2 + \bar{v}^2}{2} + \frac{\bar{p} + \bar{\gamma}\bar{P}_0}{\bar{\rho}(\bar{\gamma} - 1)} + B(\bar{\alpha}), \quad E = \frac{u^2 + v^2}{2} + \frac{p}{\rho(\gamma - 1)}, \qquad (5)$$

$$F = \frac{\alpha\bar{\alpha}}{\mu_c}(\bar{p} - \tilde{p} - \beta), \quad \beta = \bar{\alpha}\bar{\rho}\frac{dB}{d\bar{\alpha}}. \qquad (6)$$

Here $\alpha$ is the volume fraction, $u$ is the $x$-component of the velocity, $v$ is the $y$-component of the velocity, $\rho$ is the density, $p$ is the pressure, $E$ is the total specific energy, $\bar{\gamma}$ and $\bar{P}_0$ are the constant parameters in the stiffened gas equations of state of the dispersed phase, $\gamma$ is the specific heat ratio of gas, $\mu_c$ is the coefficient of compaction viscosity, $\beta(\bar{\alpha}, \bar{\rho})$ is an intergranular stress. The bar superscript is used throughout the paper to indicate dispersed phase variables.

Velocity $(\tilde{u}, \tilde{v})$ and pressure $\tilde{p}$ are interfacial variables. They are chosen as follows[8]:

$$\tilde{p} = p, \quad \tilde{u} = \bar{u}, \quad \tilde{v} = \bar{v}. \tag{7}$$

For the problem under consideration, the following parameters are chosen in the equations of state of the phases:

$$\bar{\gamma} = 2.5, \quad \bar{P}_0 = 10^7 \text{ Pa}, \quad \gamma = 1.4. \tag{8}$$

These parameters of the stiffened-gas equation of state for the particles were obtained in our previous work[24] on the basis of a parametric study of the problem of the interaction of a shock wave with a dense cloud of particles, which corresponded to the experiments[16]. They were also used in our previous work[11] devoted to the numerical simulation of a normally incident shock wave – dense particle layer interaction.

The vector **p** contains terms related to pressure relaxation. The following condition of mechanical equilibrium at the interfacial boundary is taken into account in the current work[25]:

$$\bar{p} = \tilde{p} + \beta, \tag{9}$$

$$\beta = \bar{\alpha}\bar{\rho}\frac{dB}{d\bar{\alpha}} = -\bar{\alpha}\bar{\rho} \times a \times n \times \ln\frac{1-\bar{\alpha}}{1-\bar{\alpha}_{\text{crit}}} \times \left(\frac{B(\bar{\alpha})}{a}\right)^{\frac{n-1}{n}}, \tag{10}$$

$$B(\bar{\alpha}) = \begin{cases} B_a(\bar{\alpha}), & \text{if } \bar{\alpha}_{\text{crit}} < \bar{\alpha} < 1.0, \\ 0, & \text{otherwise}, \end{cases} \tag{11}$$

$$B_a(\bar{\alpha}) = a \times [b_1(\bar{\alpha}) - b_1(\bar{\alpha}_{\text{crit}}) + b_2(\bar{\alpha})]^n, \tag{12}$$

$$b_1(\bar{\alpha}) = (1-\bar{\alpha}) \times \log(1-\bar{\alpha}), \quad b_2(\bar{\alpha}) = (1 + \log(1-\bar{\alpha}_{\text{crit}})) \times (\bar{\alpha} - \bar{\alpha}_{\text{crit}}). \tag{13}$$

Here, $B(\bar{\alpha})$ is the potential energy of compaction, $a$ and $n$ are the parameters of the compaction law, which are characteristics of the material, $\bar{\alpha}_{\text{crit}}$ is the volume fraction of the dispersed phase, upon reaching which compaction is switched on and $B(\bar{\alpha})$ becomes nonzero. The following values of the parameters in the compaction law are chosen, close to those used in our previous work[11] in which the experiment[26] was simulated:

$$a = 3 \times 10^4 \text{ J/kg}, \quad n = 1.02, \quad \bar{\alpha}_{\text{crit}} = 0.52. \tag{14}$$

The vector **s** contains source terms describing the exchange of mass, momentum, and energy between the phases. The terms $M_x$ and $M_y$ describe the interfacial momentum exchange. The term $I$ describes interfacial energy exchange. Correlations from[15] were used to describe the interphase interaction:

$$M_x = -K(\bar{u} - u), \quad M_y = -K(\bar{v} - v), \tag{15}$$

$$\mathbf{V} = (u, v), \quad \bar{\mathbf{V}} = (\bar{u}, \bar{v}), \quad \text{Re} = \frac{\rho |\bar{\mathbf{V}} - \mathbf{V}| d}{\mu_{\text{vis}}}, \tag{16}$$

$$K = \begin{cases} 0.75 C_D \dfrac{\rho \bar{\alpha} |\bar{\mathbf{V}} - \mathbf{V}|}{d \alpha^{1.65}}, & \text{if } \alpha \geq 0.8, \\ \dfrac{150 \bar{\alpha}^2 \mu_{\text{vis}}}{\alpha d^2} + 1.75 \dfrac{\rho \bar{\alpha} |\bar{\mathbf{V}} - \mathbf{V}|}{d}, & \text{if } \alpha < 0.8, \end{cases} \tag{17}$$

$$C_D = \begin{cases} \dfrac{24}{\alpha \text{Re}} [1 + 0.15(\alpha \text{Re})^{0.687}], & \text{if } \alpha \text{Re} < 10^3, \\ 0.44, & \text{if } \alpha \text{Re} \geq 10^3, \end{cases} \tag{18}$$

$$\tilde{\mathbf{V}} = (\tilde{u}, \tilde{v}), \quad \mathbf{M} = (M_x, M_y), \quad I = -(\mathbf{M}, \tilde{\mathbf{V}}). \tag{19}$$

Here $d$ is the particle's diameter, $\mu_{\text{vis}}$ is the dynamic gas viscosity coefficient.

## 3. Numerical Method

The computational algorithm is based on the Strang splitting principle. At each time step, the hyperbolic sub-step is carried out (the first stage of the algorithm), then the pressure relaxation sub-step is used (the second stage), and finally non-differential algebraic source terms that describe interfacial interaction are taken into account (the third stage).

Formulas for the second and third stages are direct generalizations of the approach from our previous work[11] to the two-dimensional case. The effect of irreversible compaction of the layer was not taken into account this time. We describe in detail the algorithm for the hyperbolic sub-step. The hyperbolic step is carried out using the HLLC method, in contrast to[11], where the Godunov method was used. When solving the system of equations (1) the first one, i.e., the compaction equation, is approximated separately. The rest of the system at the hyperbolic step is written as:

$$\mathbf{U}_t + \mathbf{F}_x(\mathbf{u}) + \mathbf{G}_y(\mathbf{u}) = \mathbf{H}(\mathbf{u}) \bar{\alpha}_x + \mathbf{I}(\mathbf{u}) \bar{\alpha}_y, \tag{20}$$

$$\mathbf{U} = \begin{bmatrix} \bar{\alpha} \bar{\rho} \\ \bar{\alpha} \bar{\rho} \bar{u} \\ \bar{\alpha} \bar{\rho} \bar{v} \\ \bar{\alpha} \bar{\rho} \bar{e} \\ \alpha \rho \\ \alpha \rho u \\ \alpha \rho v \\ \alpha \rho E \end{bmatrix}, \quad \mathbf{F} = \begin{bmatrix} \bar{\alpha} \bar{\rho} \bar{u} \\ \bar{\alpha} (\bar{\rho} \bar{u}^2 + \bar{p}) \\ \bar{\alpha} \bar{\rho} \bar{u} \bar{v} \\ \bar{\alpha} \bar{u} (\bar{\rho} \bar{e} + \bar{p}) \\ \alpha \rho u \\ \alpha (\rho u^2 + p) \\ \alpha \rho u v \\ \alpha u (\rho E + p) \end{bmatrix}, \quad \mathbf{G} = \begin{bmatrix} \bar{\alpha} \bar{\rho} \bar{v} \\ \bar{\alpha} \bar{\rho} \bar{u} \bar{v} \\ \bar{\alpha} (\bar{\rho} \bar{v}^2 + \bar{p}) \\ \bar{\alpha} \bar{v} (\bar{\rho} \bar{e} + \bar{p}) \\ \alpha \rho v \\ \alpha \rho u v \\ \alpha (\rho v^2 + p) \\ \alpha v (\rho E + p) \end{bmatrix}, \quad \mathbf{H} = \begin{bmatrix} 0 \\ \tilde{p} \\ 0 \\ \tilde{p} \tilde{u} \\ 0 \\ -\tilde{p} \\ 0 \\ -\tilde{p} \tilde{u} \end{bmatrix}, \quad \mathbf{I} = \begin{bmatrix} 0 \\ 0 \\ \tilde{p} \\ \tilde{p} \tilde{v} \\ 0 \\ 0 \\ -\tilde{p} \\ -\tilde{p} \tilde{v} \end{bmatrix}. \tag{21}$$

Here, $\bar{e} = \bar{E} - B(\bar{\alpha})$. For the numerical solution of the defining system of equations (1) the potential energy of compaction $B(\bar{\alpha})$ is moved to the right-hand side of (1) after algebraic manipulations. The corresponded term is added to the fifth component of the vector **p** in (1) and it is taken into account at the stage of pressure relaxation, as it was done in our previous paper[11].

The finite-volume approximation of (20) is written as follows:

$$\mathbf{U}_{i,j}^{h,n+1} = \mathbf{U}_{i,j}^n - \frac{\Delta t^n}{\Delta x}\left[\mathbf{F}_{i+1/2,j}^{\text{HLLC}}(\mathbf{U}_{i,j}^n, \mathbf{U}_{i+1,j}^n) - \mathbf{F}_{i-1/2,j}^{\text{HLLC}}(\mathbf{U}_{i-1,j}^n, \mathbf{U}_{i,j}^n)\right]$$
$$- \frac{\Delta t^n}{\Delta y}\left[\mathbf{G}_{i,j+1/2}^{\text{HLLC}}(\mathbf{U}_{i,j}^n, \mathbf{U}_{i,j+1}^n) - \mathbf{G}_{i,j-1/2}^{\text{HLLC}}(\mathbf{U}_{i,j-1}^n, \mathbf{U}_{i,j}^n)\right] + \mathbf{H}(\mathbf{U}_{i,j}^n)\Delta_x \bar{\alpha} \quad (22)$$
$$+ \mathbf{I}(\mathbf{U}_{i,j}^n)\Delta_y \bar{\alpha}.$$

Here, the notations are standard. The vector $\mathbf{U}_{i,j}^{h,n+1}$ denotes the conservative variable vector in the cell $(i,j)$ at the next time level $n+1$ after the hyperbolic step $h$. A uniform grid with a cell size of $\Delta x = \Delta y$ is used. A time step $\Delta t^n$ is chosen dynamically from the CFL stability condition.

Let us describe in detail the method for the calculation of the numerical flux $\mathbf{F}_{i+1/2}(\mathbf{U}_{i,j}^n, \mathbf{U}_{i,j+1}^n)$. The remaining fluxes are calculated by analogy. Parts of the numerical flux for the gaseous and dispersed phases are calculated separately:

$$\mathbf{F}_{i+1/2,j}^{\text{HLLC}}(\mathbf{U}_{i,j}^n, \mathbf{U}_{i+1,j}^n) = \begin{bmatrix} \overline{\Phi}_{i+1/2,j} \\ \Phi_{i+1/2,j} \end{bmatrix}, \quad \mathbf{U}_{i,j}^n = \begin{bmatrix} \overline{\mathbf{W}}_{i,j} \\ \mathbf{W}_{i,j} \end{bmatrix}, \quad (23)$$

$$\Phi_{i+1/2,j} = \begin{cases} \Phi_{i,j}^n, & \text{if } S_{i+1/2,j}^- \geq 0, \\ \Phi_{i,j}^n + S_{i+1/2,j}^- \times (\mathbf{Q}_*^- - \mathbf{W}_{i,j}^n), & \text{if } S_{i+1/2,j}^- < 0 \text{ and } S_{i+1/2,j}^* \geq 0, \\ \Phi_{i+1,j}^n + S_{i+1/2,j}^+ \times (\mathbf{Q}_*^+ - \mathbf{W}_{i+1,j}^n), & \text{if } S_{i+1/2,j}^* < 0 \text{ and } S_{i+1/2,j}^+ \geq 0, \\ \Phi_{i+1,j}^n, & \text{if } S_{i+1/2,j}^+ < 0, \end{cases} \quad (24)$$

$$S_{i+1/2,j}^+ = \max(u_{i,j}^n + c_{i,j}^n, u_{i+1,j}^n + c_{i+1,j}^n), \quad S_{i+1/2,j}^- = \min(u_{i,j}^n - c_{i,j}^n, u_{i+1,j}^n - c_{i+1,j}^n), \quad (25)$$

$$S_{i+1/2,j}^* = \frac{p_{i+1,j}^n - p_{i,j}^n + \rho_{i,j}^n u_{i,j}^n(S_{i+1/2,j}^- - u_{i,j}^n) - \rho_{i+1,j}^n u_{i+1,j}^n(S_{i+1/2,j}^+ - u_{i+1,j}^n)}{\rho_{i,j}^n(S_{i+1/2,j}^- - u_{i,j}^n) - \rho_{i+1,j}^n(S_{i+1/2,j}^+ - u_{i+1,j}^n)}, \quad (26)$$

$$\mathbf{Q}_*^- = \frac{\alpha_{i,j}^n \rho_{i,j}^n (S_{i+1/2,j}^- - u_{i,j}^n)}{S_{i+1/2,j}^- - S_{i+1/2,j}^*}$$
$$\times \begin{bmatrix} 1 \\ S_{i+1/2,j}^* \\ v_{i,j}^n \\ \frac{p_{i,j}^n}{\rho_{i,j}^n} + (S_{i+1/2,j}^* - u_{i,j}^n)\left(S_{i+1/2,j}^* + \frac{p_{i,j}^n}{\rho_{i,j}^n(S_{i+1/2,j}^- - u_{i,j}^n)}\right) \end{bmatrix}, \quad (27)$$

$$Q_*^+ = \frac{\alpha_{i+1,j}^n \rho_{i+1,j}^n \left( S_{i+1/2,j}^+ - u_{i+1,j}^n \right)}{S_{i+1/2,j}^+ - S_{i+1/2,j}^*}$$

$$\times \begin{bmatrix} 1 \\ S_{i+1/2,j}^* \\ v_{i+1,j}^n \\ \frac{p_{i+1,j}^n}{\rho_{i+1,j}^n} + \left( S_{i+1/2,j}^* - u_{i+1,j}^n \right) \left( S_{i+1/2,j}^* + \frac{p_{i+1,j}^n}{\rho_{i+1,j}^n \left( S_{i+1/2,j}^+ - u_{i+1,j}^n \right)} \right) \end{bmatrix}. \quad (28)$$

Here, $c$ is the speed of sound. Numerical fluxes for the dispersed phase subsystem are obtained directly from the above formulas by adding the bar superscript to all variables.

Approximation $\Delta_x \bar{\alpha}$ in (22) is performed as follows:

$$\Delta_x \bar{\alpha} = \frac{1}{\Delta x} \left( \delta_{i+1/2,j} - \delta_{i-1/2,j} \right), \quad \delta_{i+1/2,j} = \begin{cases} \bar{\alpha}_{i,j}^n, & \text{if } \bar{S}_{i+1/2,j}^* \geq 0, \\ \bar{\alpha}_{i+1,j}^n, & \text{else}. \end{cases} \quad (29)$$

Finally, the approximation of the first equation of the system (1) looks like:

$$\bar{\alpha}_{i,j}^{h,n+1} = \bar{\alpha}_{i,j}^n - \frac{\Delta t^n}{\Delta x} \left[ \bar{S}_{i+1/2,j}^* \delta_{i+1/2,j} - \bar{S}_{i-1/2,j}^* \delta_{i-1/2,j} \right]$$
$$- \frac{\Delta t^n}{\Delta y} \left[ \bar{S}_{i,j+1/2}^* \delta_{i,j+1/2} - \bar{S}_{i,j-1/2}^* \delta_{i,j-1/2} \right]. \quad (30)$$

For a problem with moving explicit interphase boundaries, after finishing the hyperbolic sub-step in a computational cell, the dispersed phase vanishing case may arise. If, after the hyperbolic sub-step, one of two situations occurs:

$$\bar{\alpha}_{i,j}^{h,n+1} < \varepsilon \quad \text{or} \quad \bar{p}_{i,j}^{h,n+1} < 0, \quad (31)$$

then, to improve the robustness of the numerical method, the following correction is made:

$$\bar{u}_{i,j}^{h,n+1} = \bar{v}_{i,j}^{h,n+1} = 0, \quad \bar{p}_{i,j}^{h,n+1} = p_{i,j}^{h,n+1}. \quad (32)$$

The value of the small parameter $\varepsilon$ in the considered problems is taken to be equal to $10^{-6}$. In a number of simulations performed, the absence of corrections according to formula (31) and (32) leads to the crushing of the computations. Problems arise in the region near the abrupt expansion of the channel at the beginning of the cavity.

To increase the accuracy of the numerical method, component-wise minmod-reconstruction[27] of conservative vectors in computational cells is carried out before the hyperbolic step. Preliminary studies showed that for the first-order accuracy method, the granular contact and the compaction wave are numerically smeared about twice as much as for the second-order methods. Besides, the specific features of distribution of particle volume fraction in the compacted region, i.e., the transition from the area of velocities non-equilibrium to velocities equilibrium (see Section 5.2), also suffer from smearing.

The HLLC method constructed in such a way is a blend of approaches from the papers[28, 29]. In contrast to later developments[9, 30, 31, 32], the methods[28, 29] are actually a straightforward generalization of the common HLLC for the Euler equations (meanwhile, the approximation of non-conservative differential terms and the equation for the volume fraction evolution are of great importance) and do not use the elements of the Riemann problem solution for the Baer-Nunziato equations. For the problems of the interaction of shock waves with a cloud or a layer of particles, the following two typical numerical issues can be distinguished. The first one is the dispersed phase vanishing case that arises at the interface between pure ambient gas and gas with particles inside the cloud or layer. Strictly speaking, the general procedure for the construction of the Riemann problem solution for the Baer-Nunziato equations should be revisited in this case[33]. The second issue is the so-called "supersonic case" in the two-phase Riemann problem[34]. It means a very high drift velocity among phases. From the Riemann problem solution point of view, it means that solid contact lies to one side of all the characteristic fields of the gas phase[33]. This situation is not a typical one in the problems of combustion wave propagation in heterogeneous explosives but can occur when a strong shock wave and a flow behind it interacts with the boundaries of clouds of layers. HLLC solver from Furfaro and Saurel[9] is the only HLLC solver among the earlier discussed ones that directly or indirectly addresses both of these numerical issues. Taking into account also that Sugiyama et al.[7] used HLLC solver[9] for the simulation of blast wave – particle layer interaction, HLLC solver[9] looks like a good option for the simulation of two-phase particulate flows with shock waves. In the next Section we consider the Rogue shock tube test and compare our results with the experimental data and simulation results from Furfaro and Saurel[9].

**4. Verification. Interaction of a Shock Wave with a Cloud of Particles.**

Experiments by Rogue et al.[16] were carried out for the clouds, which consisted of glass and nylon particles. The original paper[16] contains pressure signals from pressure gauges upstream and downstream of the cloud for the case of glass particles only. We used this data for verification purposes in our previous works[11, 24, 35]. Later, Abgrall and Saurel[36] provided pressure signals from gauges for the case of nylon particles. These pressure signals are mainly used today when the "Rogue shock tube test" is considered. It was these data, among others, that Furfaro and Saurel[9] used to test their HLLC method. We also considered the same statement with nylon particles for comparison with both experimental data[36] and simulation results[9].

A shock wave with a Mach number of $M = 1.3$ interacts with a cloud of nylon spherical particles of diameter $d = 2$ mm and the initial volume fraction $\bar{\alpha}_0 = 0.65$. The initial true density of particles is $\bar{\rho}_0 = 1050$ kg/m$^3$. The length of the computational domain is 2.8 m. The left boundary of the domain is $x = 0$. The coordinate of the left boundary of the cloud of particles is equal to $x_L = 1.39$ m, and the coordinate of the right boundary is $x_R = 1.41$ m. At the initial time moment, a shock wave is located at the point with the coordinate $x_R$ that is at the right boundary of the cloud and has moved from the right to the left. The gas pressure is recorded with the use of two transducers located at the points with the coordinates $x_1 = 1.367$ m (downstream transducer) and $x_2 = 1.520$ m (upstream transducer). At the initial time moment, the area $[0; x_R]$ is filled with the quiescent air under normal conditions. The non-penetrating condition is set at the left boundary, and the inflow condition with the parameters behind the shock wave with Mach number 1.3 is set at the right boundary. The simulation time is 4 ms and corresponds to the compression phase duration behind the incident shock wave in the experiments.

The following parameters are chosen in the equations of state of the phases:

$$\bar{\gamma} = 2.5, \quad \bar{P}_0 = 10^8 \text{ Pa}, \quad \gamma = 1.4, \tag{33}$$

and in the sub-model of intergranular stresses:

$$a = 3 \times 10^4 \text{ J/kg}, \quad n = 1.02, \quad \bar{\alpha}_{\text{crit}} = \bar{\alpha}_0 = 0.65. \tag{34}$$

The number of cells in the $x$-direction is equal to 22 400. One cell is taken in the $y$-direction to use two-dimensional code for the simulation of a one-dimensional problem. The CFL number is equal to 0.5. So, the statement is just the same as in our previous paper[11], except for the diameter and true density of particles.

Fig. 2 shows pressure signals from both pressure transducers in our simulation, in the simulation of Furfaro and Saurel[9], and in experiment[36]. The initial amplitude of the reflected wave at transducer no. 2 is reproduced well in both simulations. However, for some reasons, in the simulation of Furfaro and Saurel[9] for later points in time, the pressure at the upstream transducer systematically overshoots the experimental values in contrast to our simulation. The time of the cloud arrival at the downstream transducer is about 1.8 ms. Up to this instant, the slope of the pressure curve at the downstream transducer in the experiment is somewhat in between the calculated values. At the instant of about 2.3 ms the dashed blue line has a prominent peculiarity with no relevant physical reasons for that. In contrast to the experimental data and our simulations, for 1.8 ms $< t <$ 2.7 ms, the slope of the pressure curve[9] remains almost the same. It leads to overlarge pressure at the dropping part of the dashed curve after 2.7 ms.

In their early paper[36], Abgrall and Saurel obtained almost the same simulation results for the Rogue shock tube test as in the paper[9]. They also expected that the results could be improved by taking into account granular pressure and energy. However, as Poroshyna and Utkin[11] showed by direct comparison, the absence of granular pressure is not crucial for the Rogue test as the cloud has already been near the close packing limit at the initial time instant. All the differences in this regard take place only during the initial stage of compaction wave propagation inside the cloud, not at the stage of cloud dispersion. Apparently, significant discrepancies between the experiment and simulations could be caused by the Discrete Equations Method itself, which is in fact used in both simulations[36, 9]. Note that the Rogue test was also calculated by Sugiyama et al.[37] using the HLLC method[9]. Sugiyama et al.[37] considered the case with glass particles. Their results are free from the negative features of the simulation of Furfaro and Saurel[9] discussed above. However, they obtained a large error in the time of the main pressure rise at transducer no. 1 in comparison with the experimental data. Besides, this pressure rise is not a sharp one, but a smooth, prolonged transition region that is not relevant to the experimental data.

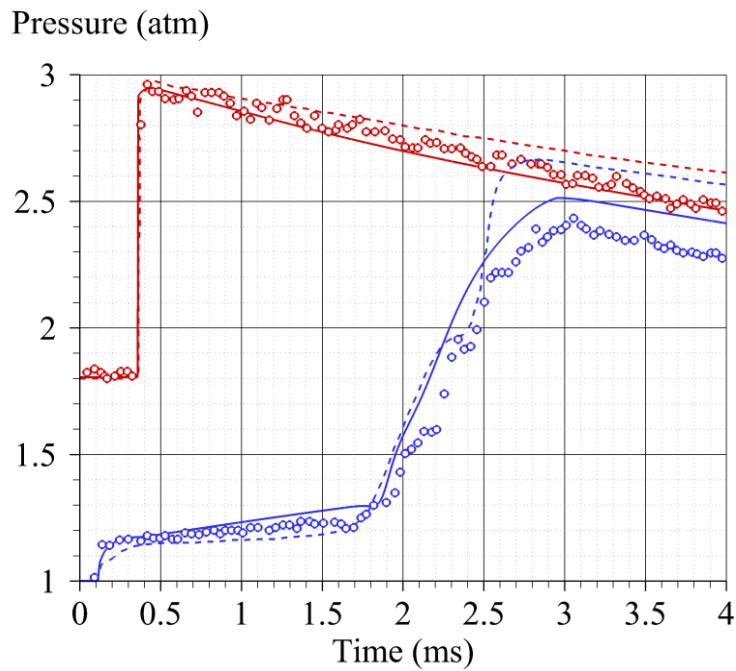

**Fig. 2.** Pressure records on the downstream transducer no. 1 (blue color) and upstream transducer no. 2 (red color). Solid lines – authors' simulation, dashed lines – simulation by Furfaro and Saurel[9] using their HLLC method, dots – experimental data from Abgrall and Saurel[36].

An additional understanding of the features of the proposed HLLC method is provided by the spatial distributions of variables, see Fig. 3. Fig. 3a shows typical distributions of the volume length of particles and pressure with the main structures of the realized flow. The shape of the distribution of the particle volume fraction is due to the interaction of the initial compaction wave in the cloud with the free boundary of the cloud; see Poroshyna and Utkin[11]. Note that during the advection of the cloud, its right border, which was initially located at the point with the coordinate $x_R$, moves in such a way that the volume fraction at $x_R$ drops to physically insignificant values, see Fig. 3b. In other words, the right boundary of the cloud moves as a localized wave, as it should be in experiments and as it is clearly visible from the experimental images in Rogue et al.[16]. It is not so in simulation in Rogue et al.[16], McGrath et al.[38], Kimura and Matsuo[39], and Poroshyna and Utkin[11]. Spatial distributions of particle volume fractions for the Rogue test are available in these papers. In contrast to the blue line in Fig. 3b, see the highlighted area, the particle volume fraction continuously increased from 0, starting from the point $x_R$ in the direction of the cloud movement.

So, firstly, the results presented in this Section show that the proposed numerical method for the solution of the Baer-Nunziato equations can be used for the simulation of dense two-phase flows with particles. Secondly, the analysis of the Rogue test solution also shows that, generally speaking, the issue of developing numerical methods for solving Baer-Nunziato equations, effective for the case of particulate two-phase flows with large enough volume fractions of particles, strong shock waves, and large volume fraction gradients, is still relevant, despite the significant progress achieved in the papers discussed in this Section.

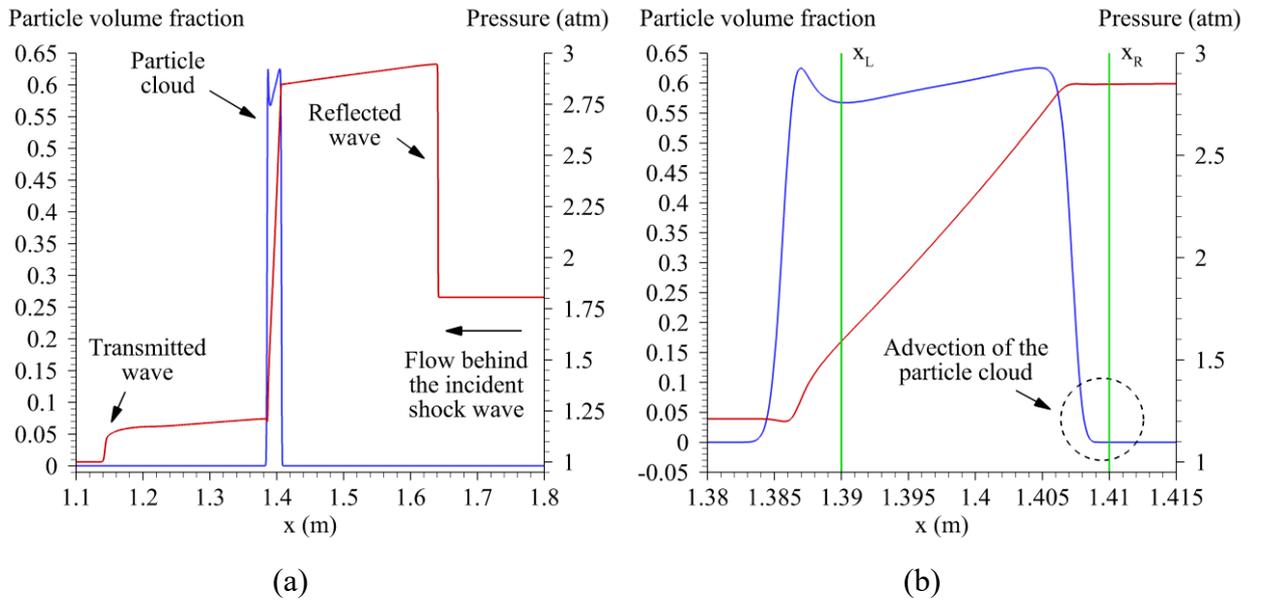

**Fig. 3.** Predicted spatial distribution of the particle volume fraction (blue line) and gas pressure (red line) for $t = 0.75$ ms: (a) general view, (b) zoomed view of particle cloud. Vertical green lines are the boundaries of the cloud at the initial instant.

## 5. Simulation Results

### 5.1. General Characteristics of the Flow Field and Procedure of the Angles Defining

A uniform grid with a cell size of $\Delta x = \Delta y = 0.25$ mm is used. The resolution is finer than that used by Houim and Oran[15] at the smallest level of adaptive mesh refinement of their numerical algorithm for the simulation of the same problem. Various intensities of the passing shock wave are considered, namely shock wave Mach numbers $M = 4.0, 3.5, 3.0,$ and $2.5$. Consider the simulation results for the shock wave with $M = 3.5$. In fact, this case corresponds to the challenging problem of the diffraction of a strong shock wave with the supersonic flow behind it around a 90° cone in a two-phase medium.

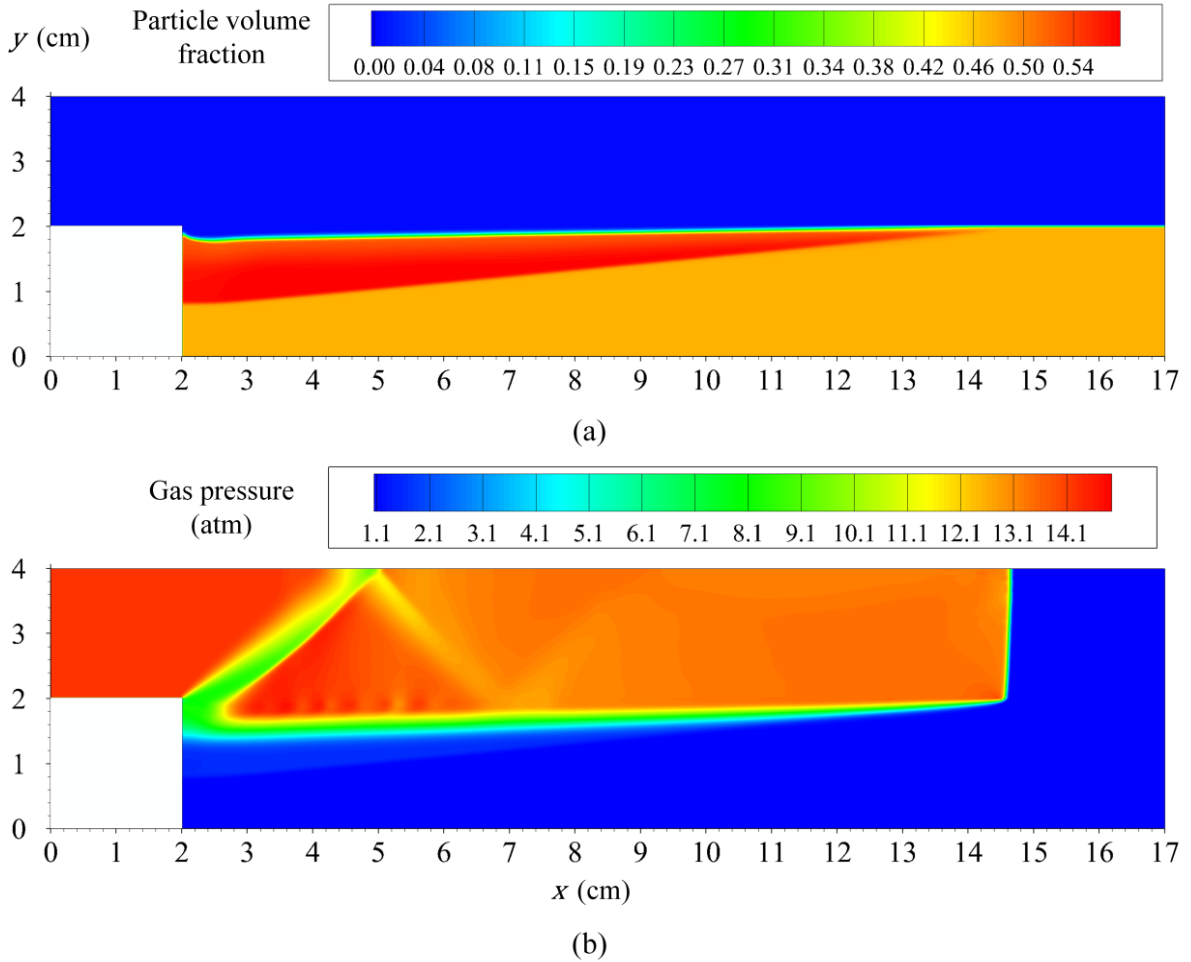

**Fig. 4.** Predicted spatial distributions of (a) particle volume fraction and (b) gas pressure. Incident shock wave Mach number is $M = 3.5$, time instant is 122 μs.

Fig. 4 illustrates the spatial distributions of the volume fraction of particles and pressure in the gas phase at a sufficiently late time instant from the beginning of the interaction process when the typical flow pattern has already been formed. Fig. 4a explains the wave pattern in the layer, and Fig. 4b explains the wave pattern in the gas phase. As it propagates, the shape of the leading shock wave is bent due to interaction with the layer. The difference in gas pressure above the layer (pressure behind the shock wave front) and inside it leads to the formation of a gas flow into the layer. Through the mechanism of interfacial momentum exchange, this initiates the movement of particles of the dispersed phase inside the layer. In turn, a local increase in the volume fraction of particles leads to the emergence of intergranular stresses that prevent excessive compaction of the layer. As the shock wave propagates, the described process leads to the formation of a region with an increased volume fraction of particles inside the layer. This area is clearly visible in Fig. 4a; it is colored red. The lower border of the red "triangle" in Fig. 4a is usually called a compaction wave (it is schematically depicted as a segment $S_2S_4$ in Fig. 1).

In experiments[12], the cavity depth was large enough not to get reflections of the compaction wave in the layer from the bottom wall of the channel. Such a statement of the experiments of Fan et al.[12] is attractive because it provides a way to study wave dynamics inside the layer of particles without additional factors that enhance the complexity of the problem. At the same time, for sure, reflection of a compaction wave from the bottom wall of the channel (taking into account its possible non-flat shape) and interaction of the reflected compaction wave with the free surface of the layer with determining the forces that act on the particles on the surface of the layer are of great importance for the consideration of the subsequent process of dust lifting[10, 17]. For the mathematical model in use, an important question will be the influence of the irreversible nature of compaction on the process. The current model should be extended by taking into account the effects of irreversible compaction, as it was done in our previous paper[11] for the normally incident shock wave – dense particle layer interaction. Consideration of cavities of smaller depth as well as cavities of more complex shapes is a matter for future studies.

The obtained flow pattern qualitatively corresponds to the experimental images[12] and the calculated results[12, 14, 15]. Starting from a certain point in time after the shock wave propagates along the layer, away from the beginning of the cavity, the flow pattern becomes self-similar. It is determined by the velocity of propagation of the shock wave along the layer and the velocity of propagation of the compaction wave inside the layer. For this reason, the solution is determined by the angles $\varphi$ and $\psi$, see Fig. 1. The obtained distributions make it possible to estimate these angles. For this purpose, we consider an arbitrary instant $t$ when the leading shock is close to the right boundary of the domain. The location of the propagating shock wave in the vicinity of the layer at that instant is denoted as $x_{SW}$. We have also chosen an arbitrary point with the abscissa $x_{1D}$ far from the propagating shock wave but also at some distance from the beginning of the cavity to eliminate the effect of flow rearrangement near the expansion of the channel. One-dimensional distribution of particle volume fraction $\bar{\alpha}$ along vertical direction at the point $x_{1D}$ is extracted from the two-dimensional distribution of $\bar{\alpha}$, see Fig. 5.

The one-dimensional plot in Fig. 5b clearly demonstrates granular contact, compaction wave, and non-uniform distribution of particle volume fraction inside the compacted part of the layer. The locations of granular contact $y_\varphi$ and compaction wave $y_\psi$ in Fig. 5 can be determined manually as the centers of the corresponding smeared profiles in Fig. 5b. The value of $y_\psi$ is the coordinate at which the particle volume fraction is equal to $(\bar{\alpha}_0 + \bar{\alpha}_{max})/2$. The value of $y_\varphi$ is the coordinate at which the volume fraction is equal to $(\bar{\alpha}_{cont} + 0)/2$. After that, the angles $\varphi$ and $\psi$ can be then estimated as:

$$\operatorname{tg} \varphi = \frac{y_{init} - y_\varphi}{x_{SW} - x_{1D}} \approx 0.0163 \to \varphi \approx 0.9°, \quad (35)$$

$$\operatorname{tg} \psi = \frac{y_{init} - y_\psi}{x_{SW} - x_{1D}} \approx 0.0996 \to \psi \approx 5.7°. \quad (36)$$

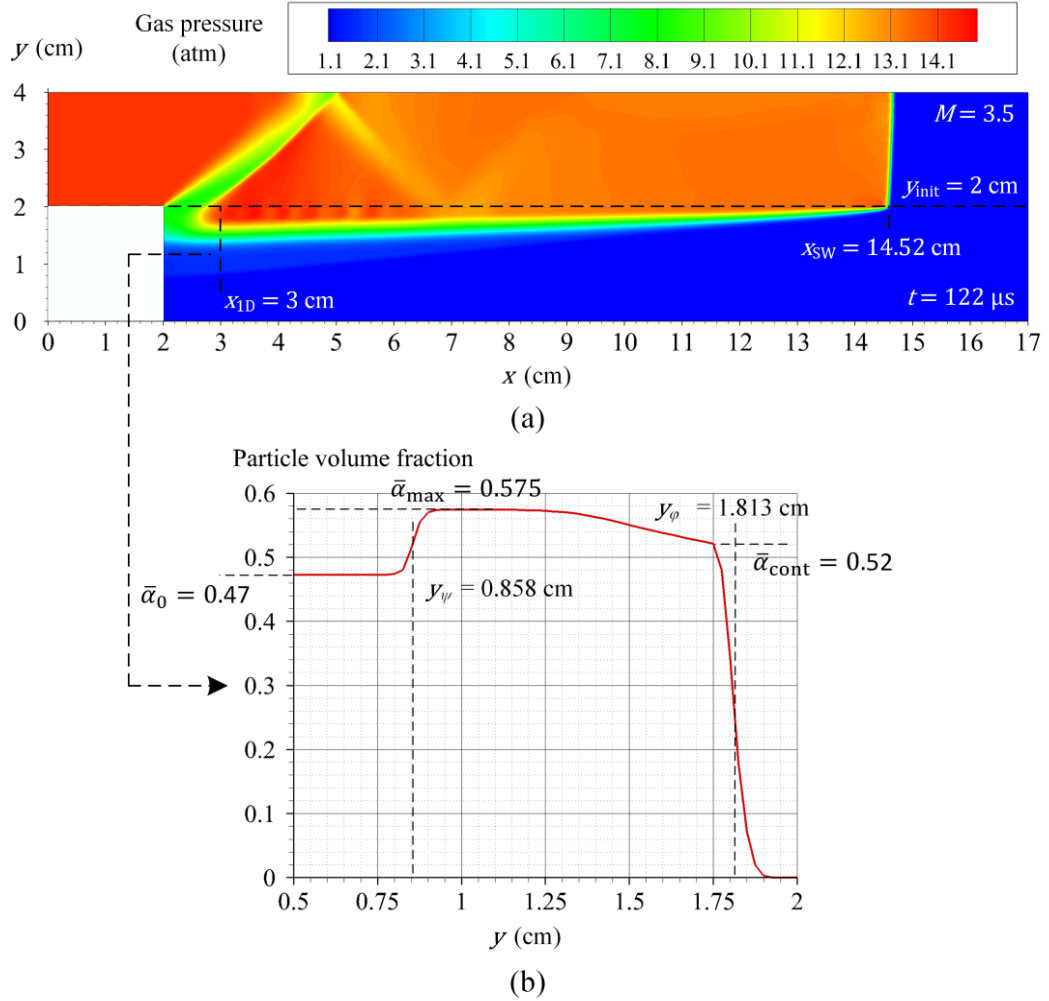

**Fig. 5.** To the procedure for defining angles $\varphi$ and $\psi$: (a) two-dimensional pressure distribution; (b) extracted one-dimensional distribution of particle volume fraction.

Tab. 1 summarizes the available data on these angles from existing experimental and numerical studies. None of the authors[12, 14, 15] explain in detail how these angles were measured. So, it is not possible to estimate the error of the data in Tab. 1, which is especially important for experimental data. Probably, understanding this, the authors of the experiments[12] gave a succinct formula, which, probably, should be taken as a reasonable characteristic of the experimental data

– "in our experiments, the measured values of (angles) are about 4° and 1–2°". In our simulation, the obtained value of the angle $\varphi$ is in good agreement with the data in Tab. 1, while the angle $\psi$ is larger than average. We verified that, for example, an increase in the parameter $\bar{\alpha}_{\text{crit}}$ in (14) from 0.52 to 0.54 leads to a decrease in the $\psi$ value at 0.3° and to an increase in the $\varphi$ value at 0.1°. That is, we obviously have a number of degrees of freedom – model parameters (8) and (14) or additional physical effects to the mathematical model, for example, interfacial heat interaction – to make the coincidence between calculated results and some averaged data in Tab. 1 better, but it's not a matter of much interest. We believe the most meaningful outcome of the experiments of Fan et al.[12] is the change in the dynamics of the process with the varying intensity of the shock wave passing over the layer. This issue will be discussed in the next Section.

| Source | Angle | Incident shock wave velocity, m/s | | | |
|---|---|---|---|---|---|
| | | 656 | 939 | 1049 | 1201 |
| | | Incident shock wave Mach number, $M$ | | | |
| | | 1.89 | 2.71 | 3.02 | 3.46 |
| Experiment by Fan et al.[12] | Transmitted compaction wave angle, $\psi°$ | 4.2 | 4.0 | 3.8 | 3.2 |
| Simulation by Fan et al.[12] | | 5.0 | 3.8 | 3.5 | 3.2 |
| Simulation by Khmel and Fedorov[14] | | 4.9 | 4.8 | 4.8 | 4.2 |
| Simulation by Houim and Oran[15] | | – | – | 4.8 | – |
| Experiment by Fan et al.[12] | Granular contact angle, $\varphi°$ | 1.2 | 1.4 | 1.6 | 1.8 |
| Simulation by Fan et al.[12] | | 1.3 | 1.6 | 1.7 | 1.9 |
| Simulation by Khmel and Fedorov[14] | | 1.3 | 1.4 | 1.6 | 1.8 |
| Simulation by Houim and Oran[15] | | – | – | 1.2 | – |
| Comments: 1. Incident shock wave velocities were taken from the plots in Fan et al.[12] According to the conditions of the experiment, the speed of sound was taken to be 347 m/s. 2. For some reasons, simulations by Khmel and Fedorov[14] were carried out for the incident shock wave Mach numbers 2.0, 3.0, 3.5, and 4.0, not for the Mach numbers from the experiment, although simulation results were compared with the experimental data from Fan et al.[12] directly. Experimental data from Fan et al.[12] are represented in Khmel and Fedorov[14] not very accurately. 3. Simulation by Houim and Oran[15] was carried out for the incident shock wave velocity of 990 m/s (we formally used the column for the leading shock speed of 1049 m/s), which doesn't match exactly to any of the experiments from Fan et al.[12] | | | | | |

**Tab. 1.** Experimental and calculated values of the angles $\varphi$ and $\psi$ by different authors.

## 5.2. Effect of the Incident Shock Wave Mach Number

For the four considered Mach numbers of the incident shock wave $M = 4.0, 3.5, 3.0,$ and $2.5$, we selected the time instants at which the positions of the leading shock wave would be almost the same and coincide with those in Fig. 4 and Fig. 5, i.e., $x_{SW} = 14.52$ cm, for a direct comparison of the dynamics of wave processes inside the layer, see Fig. 6. For each case, the procedure for defining angles $\varphi$ and $\psi$ from the previous Section was applied, see Tab. 2, the first and the second row. One-dimensional plots in Fig. 6 clearly show that the angle $\varphi$ increases with the increase in the incident shock wave Mach number while the angle $\psi$ decreases. The dependencies of angles $\varphi$ and $\psi$ from $M$ are close to linear ones, as can be seen from Tab. 2.

Let's give an explanation of the result. The ratio between the velocity of the leading shock, the velocity of the lowering of the granular contact $v_{gc}$, and the velocity of the compaction wave $v_{gc}$ determines the angles. Tab. 2 contains estimations for $v_{gc}$ and $v_{cw}$ in the vertical direction, as well as the integral momentum of the gas pressure force that acts on the layer in each case. The momentum is estimated as follows: the pressure $p_M$ behind the shock wave with a given Mach number $M$ multiplies by the time interval $\tau$ of the shock wave propagation from the beginning of the channel expansion to the point $x_{SW}$. For a stronger shock wave, the duration of the impact of the shock wave on the layer is less, but the pressure exerted on the surface is greater. As can be seen in Tab. 2, for a stronger wave, the net pressure force momentum acting on the layer is greater than for a weak one. This can explain the dynamics of the change in $\varphi$ angle; that is, for a stronger shock wave, the surface of the layer lowers more.

Velocities $v_{gc}$ and $v_{cw}$ are estimated as (see Fig. 5):

$$v_{gc} = \frac{y_{init} - y_\varphi}{\tau}, \quad v_{cw} = \frac{y_{init} - y_\psi}{\tau}. \tag{37}$$

Note that the dynamics of $v_{gc}$ and $v_{cw}$ changes depending on $M$ are qualitatively the same, see Tab. 2, the last two rows. For this reason, it could be assumed that the momentum of the gas pressure force transmitted to the layer will lead to a change in the $\psi$ angle, the same as the $\varphi$ angle. However, this does not happen. The reason for that is the difference in mechanisms of propagation between the granular contact and the compaction wave inside the layer. Spatial distributions of defining parameters inside the layer are necessary to explain this difference.

For clarity, we compare the cases corresponding to the strongest and weakest incident shock waves among the ones considered. Fig. 7 shows one-dimensional distributions of the defining parameters across the layer. High pressure behind the propagating shock wave produces a gaseous pressure wave propagating inside the layer.

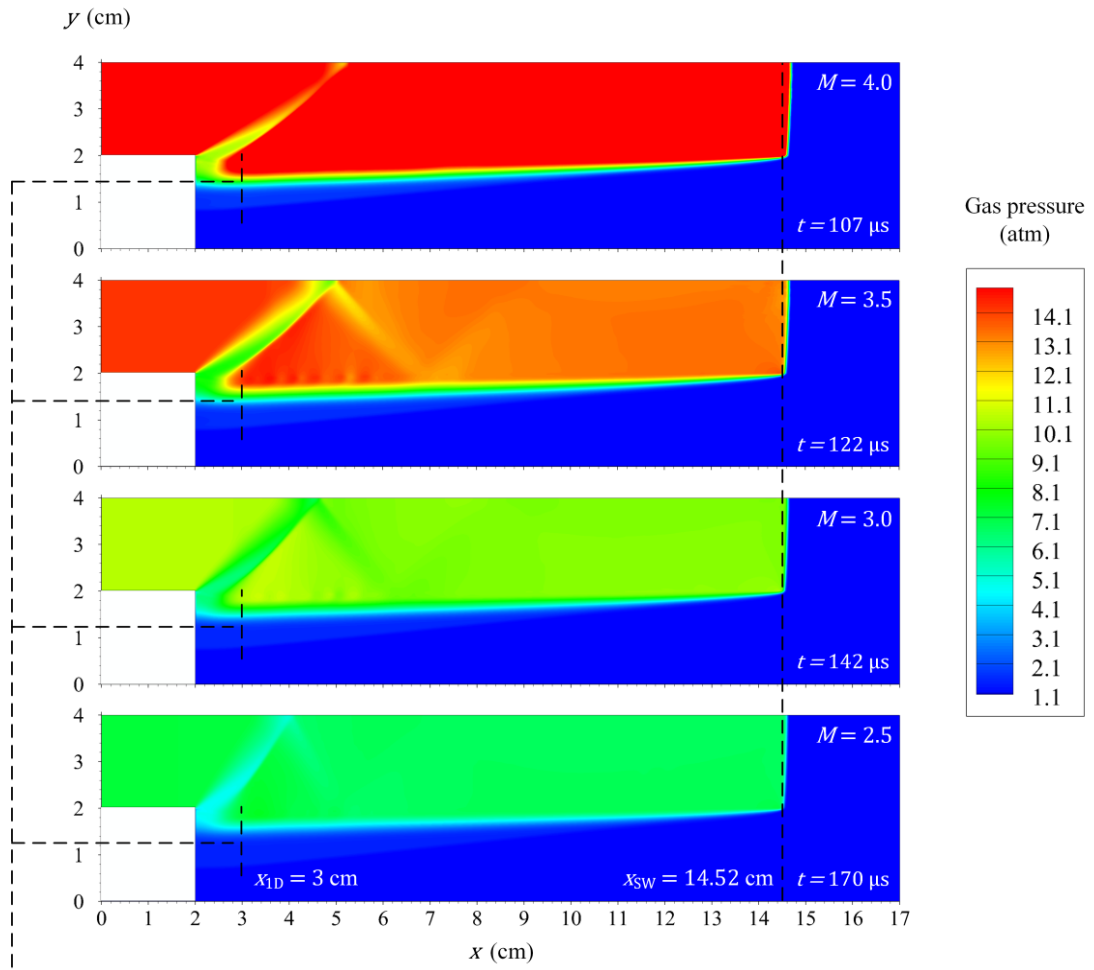

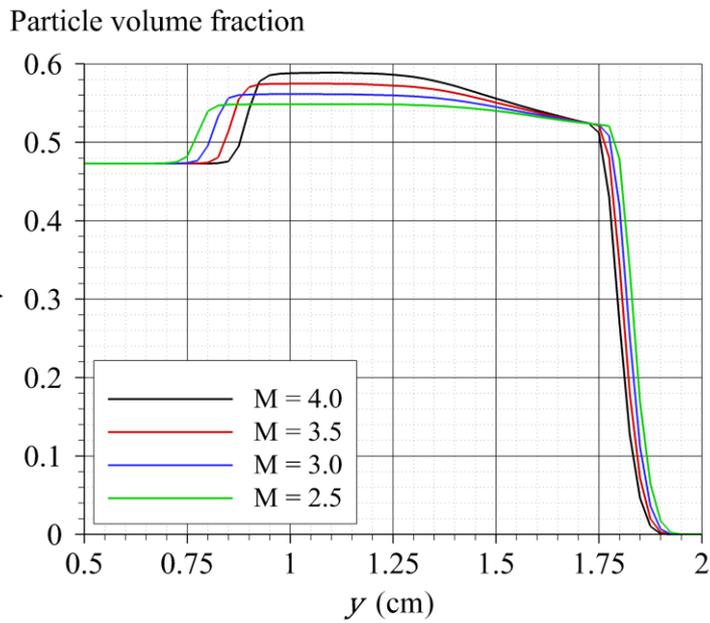

**Fig. 6.** The effect of the incident shock wave Mach number on wave processes in the layer: (a) two-dimensional pressure distribution for different incident shock wave Mach numbers; (b) corresponding extracted one-dimensional distributions of particle volume fraction.

| Parameter | Incident shock wave Mach number, $M$ | | | |
|---|---|---|---|---|
| | 4.0 | 3.5 | 3.0 | 2.5 |
| Transmitted compaction wave angle, $\psi°$ | 5.49 | 5.69 | 5.90 | 6.10 |
| Granular contact angle, $\varphi°$ | 0.98 | 0.93 | 0.88 | 0.81 |
| Maximum particle volume fraction in the compacted part of the layer, $\bar{\alpha}_{max}$ | 0.588 | 0.575 | 0.561 | 0.548 |
| Time interval of a shock wave propagation, $\tau$, μs | 92 | 106 | 123 | 147 |
| Post-shock pressure, $p_M$, atm | 18.6 | 14.2 | 10.4 | 7.2 |
| $p_M \cdot \tau$, atm·μs | 1711 | 1505 | 1279 | 1058 |
| $v_{gc}$, m/s | 22 | 18 | 14 | 11 |
| $v_{cw}$, m/s | 120 | 108 | 96 | 83 |

**Tab. 2.** Integral parameters of the process of shock wave propagation over a dense particle layer.

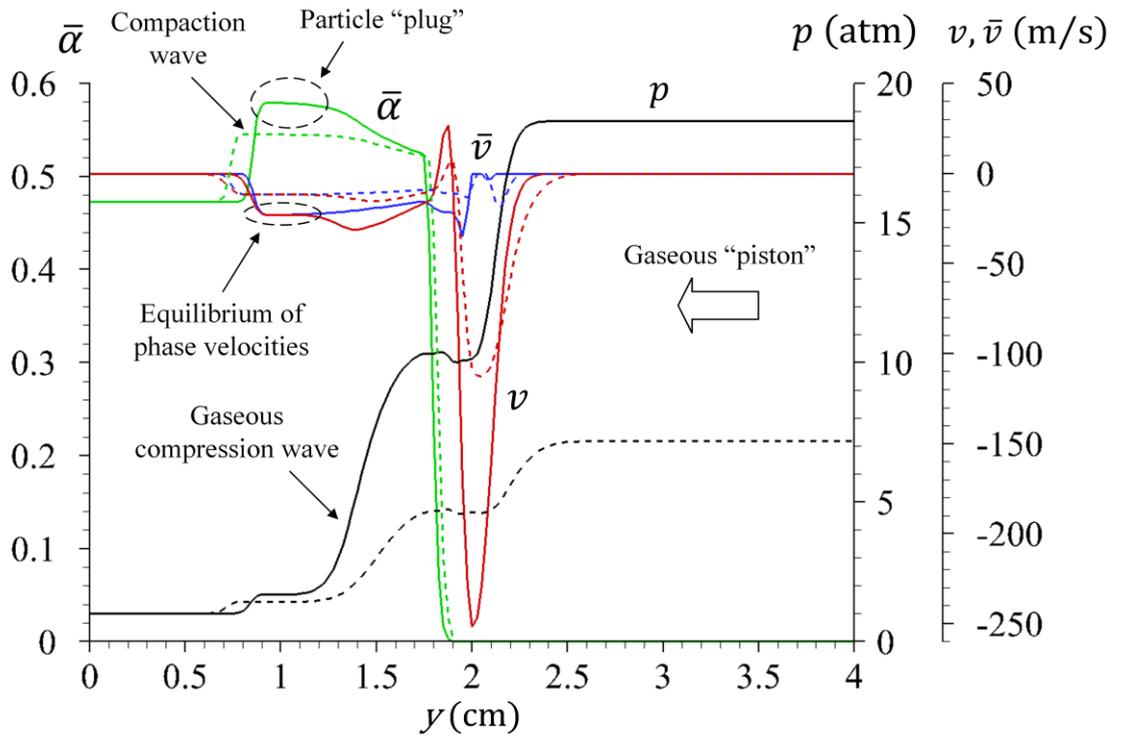

**Fig. 7.** Predicted one-dimensional distributions of particle volume fraction (green lines), vertical component of particle velocity (blue lines), vertical component of gas velocity (red lines), and gas pressure (black lines) across the channel width at $x_{1D} = 3$ cm. Solid lines correspond to the case of $M = 4.0$ (time instant 107 μs), dashed lines – $M = 2.5$ (time instant 170 μs).

In turn, this pressure wave causes particles in the layer to move. A comparison of the blue and red curves inside the layer in the vicinity of the layer surface shows nonequilibrium in the velocities of gas and particles. The particles accelerate after the gas penetrates the layer. As we move deeper into the layer, the velocities of the gas and particles reach equilibrium. At this moment, the volume fraction of particles, which had previously increased deep into the layer, reaches an almost constant value, depending on the intensity of the shock wave propagating over the layer. A "plug" is formed, the leading edge of which is called the compaction wave. So, the compaction wave propagates ahead of the compression gas wave in the layer. The main elements of the mechanism of propagation of disturbances in the layer are marked in Fig. 7 for curves corresponding to $M = 4.0$.

A stronger shock wave above the layer provides a stronger compression gas wave in the layer and faster particle motion. This factor promotes the compaction wave velocity. However, the stronger compression wave provides an increase in the volume fraction of particles in the layer to larger values. The increased volume fraction of particles in the plug makes it difficult for the gas to move further into the layer. This factor works in the opposite direction. So, the mechanism of particle plug formation is a limiting factor for the speed of disturbances propagating in the phase of particles inside the layer. It leads to the inverse dependence of the $\psi$ angle on the shock wave Mach number in comparison with the $\varphi$ angle.

It is important to note that the described mechanism of wave propagation in the layer of particles is qualitatively very similar to the mechanism of convective burning in an extended charge of heterogeneous explosive. In our previous paper[40], combustion wave propagation in a black powder charge was simulated using the Baer-Nunziato-like model. The setting of the problem from the work of Ermolaev et al.[41] was used. Due to the weak dependence of the burning rate of the black powder on the pressure, the combustion wave in the charge propagates, as a rule, almost at a constant speed without transitioning to detonation. This is ensured by the formation of a similar plug with an increased volume fraction of particles due to a compaction wave propagating in front of the heat release zone. The plug prevents the gas from penetrating forward, and the velocity of propagation of disturbances along the charge is ultimately determined by the velocity of propagation of the compaction wave. The described mechanism was obtained in the paper[40] within the framework of the Baer-Nunziato model, similar to the one used in this work. In our case, there is no burning of particles. The gas pressure behind the shock wave above the layer permanently acts as a piston instead.

## 6. Conclusions

1.  The paper demonstrates that wave processes in a dense layer of particles over which a shock wave propagates can be described qualitatively and quantitatively correctly using the two-fluid Baer-Nunziato model, taking into account intergranular stresses in the phase of particles. The problem under consideration corresponds approximately to full-scale experiments[12]. In a similar setting, the adequacy of such a mathematical model was also demonstrated in the paper[7]. Note that, generally speaking, the Baer-Nunziato model is justified only for dense layers when the particles are in constant contact with each other. For correct modeling of problems in which the volume fraction of particles can vary widely, hybrid models are built, as, for example, in Ref. 42.

2.  A two-dimensional version of the HLLC method for the Baer-Nunziato equations based on the ingredients from the works[28, 29] is proposed. It is shown that the numerical algorithm based on the proposed HLLC method and pressure relaxation procedure, taking into account intergranular stresses in the phase of particles, is workable for problems with explicit interfacial boundaries and strong shock waves, namely, for the simulation of the shock wave propagation over the layer of particles and for the simulation of shock wave – particle cloud interaction.

3.  In simulations, reasonable agreement with the experiment is obtained for the angles of the transmitted compaction wave and granular contact from a shock wave propagating along the particle layer. Obtaining these results does not require a significant change in the parameters of the equation of state of the dispersed phase and intergranular stresses (by several times, but not by an order of magnitude), in comparison with our previous works[11, 24, 35], in which experiments[16, 26] were modelled. We attribute this to the hyperbolicity of the Baer-Nunziato model, which provides the intuitively expected continuous dependence of the properties of solutions on the a priori, very imprecisely known parameters of the model.

4.  The correct dependence of the transmitted compaction wave and granular contact angles on the intensity of the shock wave is obtained. The transmitted compaction wave angle decreases, while the granular contact angle increases with the increase in the propagating shock wave Mach number. It is found that with an increase in the intensity of the shock wave, the volume fraction of particles in the compacted region increases, which prevents the penetration of gas into the compressed particle layer. This causes the size of the compacted particle layer to decrease with an increase in the intensity of the shock wave.

**ACKNOWLEDGMENTS**

P. Utkin expresses deep appreciation to his former Ph.D. student, Ya.E. Poroshyna, for her contribution to the development of the numerical algorithm (a generalization of the pressure relaxation algorithm to the two-dimensional case) used in the study. The work of P. Chuprov was carried out under the state task of the Institute for Computer Aided Design of the Russian Academy of Sciences (project no. AAAA-A19-119041590048-0).

## AUTHOR DECLARATIONS

### Conflict of Interest

The authors have no conflicts to disclose.

### Author Contributions

**Pavel Utkin:** Conceptualization (lead); Supervision (lead); Writing – original draft (lead); Writing – review & editing (lead); Methodology (equal); Investigation (equal); Formal analysis (equal). **Petr Chuprov:** Data curation (lead); Methodology (equal); Investigation (equal); Formal analysis (equal); Writing – original draft (supporting); Writing – review & editing (supporting).

### Data availability

The data that support the findings of this study are available from the corresponding author upon reasonable request.